\documentclass{revtex4}
\usepackage{graphicx} 
\usepackage{amsmath}

\begin{document}
\title{Revisiting the Second Vassiliev (In)variant for Polymer Knots}
\author{Alexander R. Klotz and Benjamin Estabrooks}
\affiliation{Department of Physics and Astronomy, California State University, Long Beach}
\begin{abstract}
Knots in open strands such as ropes, fibers, and polymers, cannot typically be described in the language of knot theory, which characterizes only closed curves in space. Simulations of open knotted polymer chains, often parameterized to DNA, typically perform a closure operation and calculate the Alexander polynomial to assign a knot topology. This is limited in scenarios where the topology is less well-defined, for example when the chain is in the process of untying or is strongly confined. Here, we use a discretized version of the Second Vassiliev Invariant for open chains to analyze Langevin Dynamics simulations of untying and strongly confined polymer chains. We demonstrate that the Vassiliev parameter can accurately and efficiently characterize the knotted state of polymers, providing additional information not captured by a single-closure Alexander calculation. We discuss its relative strengths and weaknesses compared to standard techniques, and argue that it is a useful and powerful tool for analyzing polymer knot simulations.
\end{abstract}

\maketitle

\section{Introduction}

In this work, we examine the use of the Second Vassiliev Invariant ($v_2$) \cite{vassiliev1990cohomology} for the classification of open polymer knots, in comparison to the most commonly used tool of evaluating the Alexander polynomial ($\Delta(t)$) after virtually closing the knot \cite{alexander1928topological}. While mathematical knots are only defined in closed loops, colloquially and in technological use any entangled portion of an open fiber may be identified as a knot, as anyone who wears shoes or headphones can attest. Indeed, most human interactions with knots involve tying or untying them in open strings or ropes. Inspired in part by the occurrence of knots in viral and cellular DNA \cite{arsuaga2002knotting,siebert2017there}, experiments examining the polymer physics of knotted DNA molecules \cite{klotz2017dynamics,bao2003behavior}, and the potential impact of knots on genomics devices \cite{reifenberger2015topological, plesa2016direct}, many simulation studies have been performed examining the knotting and unknotting of open DNA-like polymer chains \cite{tubiana2013spontaneous, jain2017simulations, soh2019conformational, rieger2016monte, caraglio2019topological}. 

Several schemes exist to extend the definition of knots to open curves. The most common is closure, in which the two ends of the chain are connected to form a closed loop, such that the knot can be identified through the calculation of an invariant such as the Alexander polynomial \cite{alexander1928topological}. The ends may be connected directly through the interior of the knot, or connected to a virtual surface outside the knot. A popular and freely available tool, KymoKnot, performs ``minimally interfering closure'' either by connecting the ends directly or via the convex hull of the knot, depending on whether the ends are closer to each other or the surface \cite{tubiana2018kymoknot}. While these methods yield a definite topological classification (for simple knots that do not share an Alexander polynomial), open curves do not necessarily have a definite knot topology. More rigorously, the ends of chain may be projected to many points on a virtual sphere surrounding the knot, computing a distribution of possible topologies consistent with the curve \cite{alexander2018projecting,millett2013identifying} (Figure 1a). First discussed in the context of protein knotting \cite{mansfield}, this is sometimes known as stochastic closure, although the points need not be chosen randomly. Although this better captures the ambiguous topology of an open knot, it is significantly slower computationally. Besides closure, other methods include the classification of knottoids \cite{turaev2012knotoids}, which categorize incomplete closures, and virtual knots, which categorize ambiguous closures \cite{kauffman2021virtual}.

To demonstrate the utility of the $v_2$, we focus on two cases in which open knotted polymers may have ambiguous topology. The first is the untying of complex knotted chains, in which knots evolve through a sequence of progressively simpler topologies until they reach the unknot (Figure 1b). The second is the case of spherically confined polymers, in which the motion of the ends of the chain through the polymer-filled sphere leads to a diffusive equilibrium of knotting and unknotting. Alexander \cite{alexander2018projecting} defines ``strong'' and ``weak'' knotting based on the stochastic closure of knots to the surface of a bounding sphere, defining strong knotting when more than half the projections are represented by a single knot type, weak knotting when there is only a plurality, and unknottedness when over half the projections yield the unknot. An unconfined polymer knot is typically strongly knotted when the chain ends are far from the knotted core, but can become weakly knotted as the ends of the chain reptate through the knot as it unties. In tightly confined polymers in spheres (which mimics the packaging of viral DNA in capsids \cite{arsuaga2002knotting}), the ends of the chain often cannot be connected or projected to the surface of the sphere without creating several new crossings. Not only is the knotting weak, it is difficult to unambiguously define what the topology ``would be'' if the ends were connected. Dai and Doyle \cite{dai2018universal} examined the complexity of highly confined knots, and broke the ambiguity by imposing an arbitrary axis to stretch the chain along, physically rather than virtually realizing the topology.

For experimental context, studies by one of the authors of this manuscript examined knotted DNA molecules with topologies that were indeterminate, variable, and weak. To form knots in DNA, molecules are compressed through an electrohydrodynamic instability until they are much smaller than their equilibrium size \cite{tang2011compression, radhakrishnan2021collapse}, forming dense weak knots similar to those found in polymers under extreme confinement. When these compressed DNA molecules are stretched and examined with fluorescence microscopy, they typically have complex knots localized in the interiors of the molecules \cite{renner2015stretching}. When these knots reach the end of the chain, they form a diffraction-limited spot within which the end reptates through the knot, gradually untying \cite{soh2018untying}. Often, the knot only partially unties and separates from the chain end, participating in multiple untying events until the molecule no longer contains a knot \cite{narsimhan2017steady}. To understand this process, we carried out simulation studies with twist knots that could untie in multiple stages \cite{soh2019conformational}, similar to the torus knots studied in the current manuscript. In both scenarios, knot formation and partial knot untying, it is desirable to have a metric of the molecule's topology that doesn't just characterize the topology before and after, but during these transitions.



\begin{figure}
    \centering
    \includegraphics[width=\textwidth]{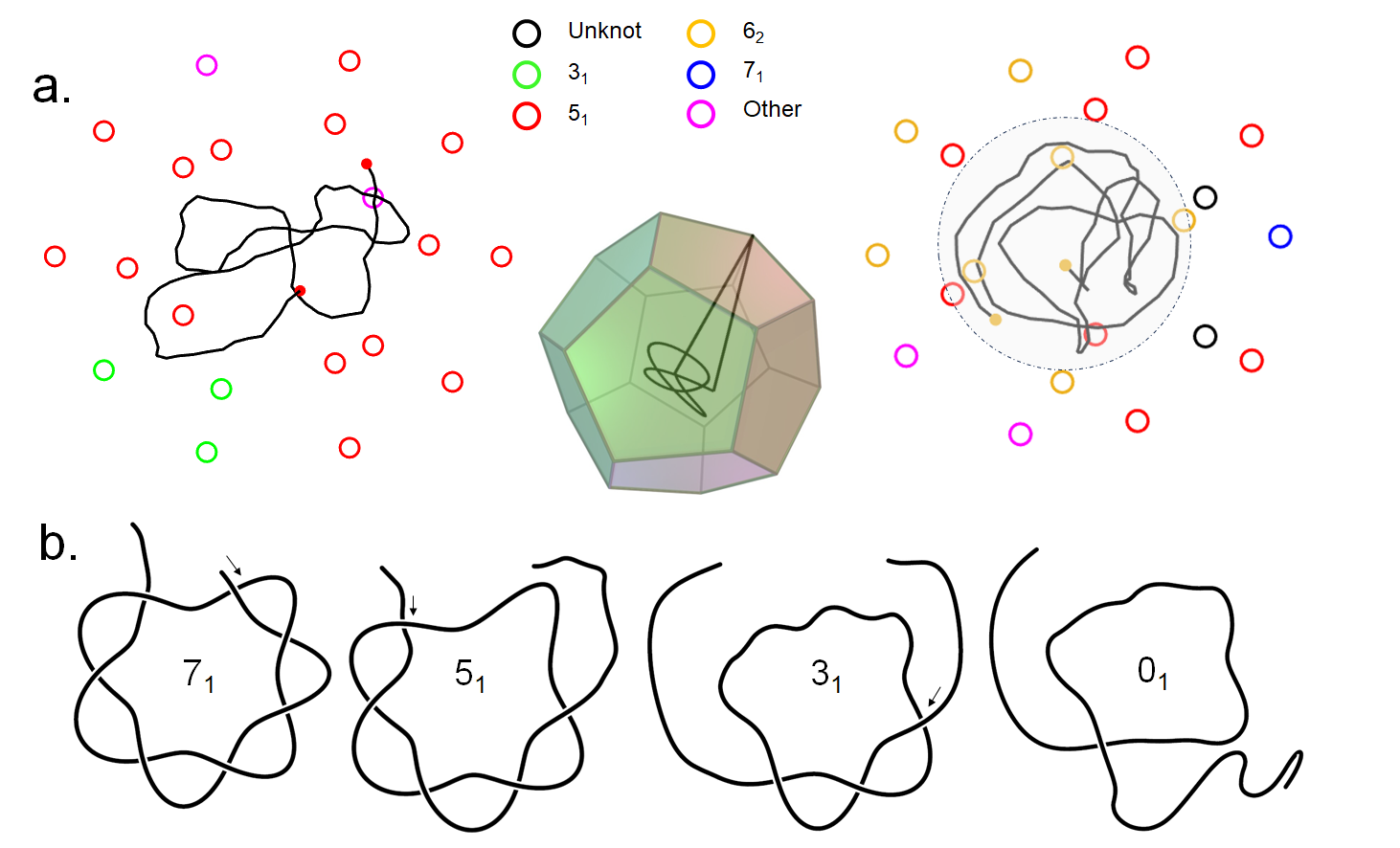}
    \caption{a. Open knotted polymers and their knot types as determined by closure. The open circles surrounding each chain represent the knot type as determined by connecting both ends to the 20 corners of a dodecahedron. The filled circles on the chain ends represent the knot type determined by connecting them directly. Left: an unconfined polymer untying between a strong $5_1$ and $3_1$ topology, with 15/20 projected closures and the end closure yielding $5_1$. Right: A confined polymer that is weakly knotted, with 8/20 projected closures yielding $5_1$ and 7/20 projected closures plus the end closure yielding $6_2$. The radius of the closure points was much larger than in the visualizations. Midset shows an open trefoil knot closed by projection to one corner of a dodecahedron. b. The typical untying sequence simulated in this manuscript, in which a $7_1$ unties into a $5_1$ then into a $3_1$ and finally into the unknot.}
    \label{fig:proj}
\end{figure}

To measure the topological state of a knotted polymer chain we desire a metric with the following properties:
\begin{itemize}
    \item it should take Cartesian coordinates as an input.
    \item for closed chains, it should reproduce values of established knot invariants.
    \item for open chains, it should continuously interpolate between these established values as the chain ends move with respect to the rest of chain, particularly when untying or forming a new knot.
    \item it should identify subsections of the chain that contribute most significantly to its topological state.
    \item it should be computable in a reasonable amount of time.
\end{itemize}

The  closure-Alexander technique used by KymoKnot satisfies four of these requirements for strongly knotted curves, but does not provide a continuous measurement for open chains. When untying events occur, the measured Alexander polymer suddenly shifts, and can fluctuate back and forth repeatedly as the knot unties. Alexander determination with stochastic closure can satisfy the third requirement, at the expense of computation speed. Other open knot classification schemes exist but do not satisfy all requirements. Sleiman et al. \cite{sleiman2024geometric} showed a neural network trained on writhe representations of knots can track a knot's topology as it unties, but it cannot be identified with known invariants and lacks smooth interpolation. 


The Second Vassiliev Invariant ($v_2$) is a parameter that characterizes the linking of a curve with itself \cite{vassiliev1990cohomology}. It is similar to the space writhe of a knot, with the additional complexity that it characterizes the linking of two identical curves with each other as measured over alternating indices. It is equivalent to the quadratic coefficient of the Alexander-Conway polynomial \cite{smith2022second}, and related to the Casson invariant \cite{zhao2012numerical}. While it has not seen as much use as the Alexander polynomial, there has been some use of $v_2$ in the study of polymer knots. Ferrari and Zhao \cite{zhao2014ring} as well as Daeguchi and Tsurusaki \cite{deguchi1994statistical} have used a version of the Vassiliev invariant to categorize knotting in closed knotted curves. In two works, Moore et al. \cite{moore2004topologically} and Lua et al. \cite{lua2004fractal} used it in conjunction with the Alexander Polynomial to identify knotting probabilities in random polygons. There has been similar limited use of the Casson invariants to chararacterize random petal knots \cite{even2017models}. 


Recently, Panagiotou and Kauffman established a definition of $v_2$ for open chains \cite{panagiotou2021vassiliev}, which was further used to measure $v_2$ for random walks \cite{smith2022second}. The purpose of this manuscript is to show that a discrete version of the open-curve $v_2$ satisfies all five requirements for a good categorizer of open knots, based on an analysis of Langevin Dynamics simulations of knotted polymers. We will compare its utility to that of the Alexander polynomial, discussing its relative advantages and disadvantages. Before proceeding, we note that $v_2$ is not as strong an invariant as the Alexander polynomial. For example, the $3_1$ and $6_3$ knots have the same value of $v_2$, while having different Alexander polynomials. We also note that the topology of an open curve that evolves over time is not fixed, meaning parameters calculated from its coordinates are not invariant. We refer to the value we compute as the \textbf{Vassiliev parameter}.




\section{Theory and Simulations}

\subsection{The Vassiliev Parameter}

The Second Vassiliev Invariant can be calculated from the double alternating self-linking integral (SLL) defined by Panagioutou and Kaufmann \cite{panagiotou2021vassiliev}. A curve in space takes the form $\vec{r}$(s) where s is a parametric variable between 0 and 1 and its tangent vector as $\dot{\vec{r}}$(s). If we consider four variables traversing the curve between 0 and 1, $s_{1}>s_{2}>s_{3}>s_{4}$, the double alternating self linking integral can be defined as:

\begin{equation}
    SLL=\frac{1}{8\pi}\int_{0}^{1}\int_{0}^{1}\int_{0}^{1}\int_{0}^{1} 
   \left( \left(\dot{\vec{r}}(s_{1})\times\dot{\vec{r}}(s_{3}) \right)\cdot\frac{\vec{r}(s_{1})-\vec{r}(s_{3})}{|\vec{r}(s_{1})-\vec{r}(s_{3})|^{3}}\right)   \left( \left(\dot{\vec{r}}(s_{2})\times\dot{\vec{r}}(s_{4}) \right)\cdot\frac{\vec{r}(s_{2})-\vec{r}(s_{4})}{|\vec{r}(s_{2})-\vec{r}(s_{4})|^{3}}\right)
 ds_{4}ds_{3}ds_{2}ds_{1}
\end{equation}

We can compute this for a knot based on a discrete set of Cartesian coordinates $\vec{r}_{i}(x,y,z)$ and its tangent vectors $\vec{\dot{r}}_{i}(x,y,z)$. Instead of integrating along the path of a knot parameterized between 0 and 1, we sum over the indices of each vertex in the knot between 1 and N. We first compute the linking matrix M:

\begin{equation}
    M_{ij}=\left(\dot{r}_{i} \times \dot{r}_{j}\right) \cdot \frac{\vec{r}_{i}-\vec{r}_{j}}{ |\vec{r}_{i}-\vec{r}_{j}|^{3}}
    \label{eq:matrix}
\end{equation}

The sum of all elements in the linking matrix is 4$\pi$ times the space writhe of the knot. The double alternating self-linking integral is a sum of every unique product of elements in this matrix with alternating indices.

\begin{equation}
    SLL=\frac{1}{8\pi}\sum_{i=4}^{N}\sum_{j=3}^{i-1}\sum_{k=2}^{j-1}\sum_{\ell=1}^{k-1} M_{ik}M_{j\ell}
\end{equation}

When computed from a set of coordinates, SLL is six times the the Second Vassiliev Invariant, under the convention that $v_{2}=0$ for the unknot $v_{2}=1$ for the trefoil. Other tabulations use a different convention, e.g. Ferrari and Zhao use $v_{2}=\frac{-1}{12}$ and $v_{2}=\frac{23}{12}$ for the unknot and trefoil \cite{zhao2014ring}. To maintain consistency with the former tabulation, and to take into account that the measured quantity for an open knot is not truly invariant, we define the Vassiliev parameter:

\begin{equation}
    V=\frac{SLL}{6},
\end{equation}
which is the parameter used to analyze polymer knots in this work.

In practice, the tangent vector is computed as the average of the displacement vector between each vertex and its two neighbors:

\begin{equation}
    \dot{\vec{r}}_{i}=\frac{1}{2}(\vec{r}_{i+1}-\vec{r}_{i-1})
\end{equation}
When a chain is closed, the first tangent is calculated between the 2nd and Nth vertices, and the Nth as between the 1st and N-1th. When the chain is open, the tangent vector at either end is calcualted from the displacement vector between the second and first vertex, and between the final and penultimate vertex. In practice, the tangent vector could be computed more smoothly e.g. by fitting a spline curve to each quartet of points and computing its derivative.

Other methods of integration may produce more accurate approximations of the integral in Equation 1. Ferrari and Zhao, for example, used Monte Carlo integration while smoothing the corners of the curve near vertices \cite{ferrari2014monte}. They were able to get a very precise measurement of the Vassiliev invariant of a 24-vertex cubic lattice trefoil knot using one billion Monte Carlo samples, in contrast to the 13824 multiplications that Eq. 3 requires. In this work we demonstrate the utility of our coarse-grained integration scheme, knowing that it can likely be further improved.



\subsection{Langevin Dynamics Simulation}

We simulate polymer knots using a model used to simulate topologically complex polymers in previous works \cite{sleiman2024geometric,caraglio2019topological}. We use a parameterization that models DNA at the low ionic strengths used in fluorescence experiments, in which the persistence length is ten times the effective width of the molecule \cite{tree2013dna}, and the contour length is ten to twenty persistence lengths.

In short, the chain is comprised of N beads of diameter $\sigma$ connected by springs to their two neighbors. A finitely-extensible nonlinear elastic (FENE) spring potential with a maximum extension of 1.5$\sigma$ is used. Excluded volume interactions between beads are enforced by a truncated Lennard-Jones repulsive potential that applies when the centers of mass of two beads are closer than $\sigma$. The relatively short range of distances between the excluded volume of the beads and maximum extension of the springs ensures that strands do not cross and the knots cannot untie themselves by pathological strand crossings. Bending rigidity is imposed by a Kratky-Porod potential depending on the cosine of the angle between three successive beads. The strength of this potential sets the persistence length of the polymer. The entire contribution to the energy of a bead is:

\begin{equation}
    U_{tot}=U_{spr}+U_{ev}+U_{bend}+U_{conf}.
\end{equation}
The excluded volume interaction takes the form:
\begin{equation}
    U_{ev}=\begin{cases}
  4\epsilon\left[\left(\frac{\sigma}{r}\right)^{12}-\left(\frac{\sigma}{r}\right)^{6}+\frac{1}{4}\right]
    & \text{if } r\leq 2^{1/6}\sigma\\
    0              & \text{otherwise},
\end{cases}
\end{equation}
where $\epsilon$ sets the energy scale of the repulsive interactions. If activated, the confinement energy takes the same form as the excluded volume interaction, activating when r is within $2^{1/6}\sigma$ of the sphere's radius R. The spring force is parameterized as:
\begin{equation}
    U_{spr}=\begin{cases}
    -\frac{1}{2}(\kappa\frac{\epsilon}{\sigma^2})R_{max}\log(1-(\frac{r}{R_{max}})^{2}),  & \text{if } r\leq R_{max}\\
    0             & \text{otherwise},
\end{cases}
\end{equation}
where $\kappa$ is typically 30 and sets the spring constant in units of $\epsilon/\sigma^2$ and $R_{max}$ is the maximum separation of the springs, and is 1.5$\sigma$ in this work. The bending potential takes the form:
\begin{equation}
    U_{bend}=\frac{\ell_{p}}{\sigma}kT(1-\cos{\theta}).
\end{equation}
The dimensionless ratio of the persistence length $\ell_{p}$ to the bead diameter is typically 10 in this work. The time evolution of the ith bead is determined by the Langevin equation:

\begin{equation}
    m\ddot{\vec{r_{i}}}=-\gamma\dot\vec{r_{i}}-\nabla U_{tot}+\sqrt{2kT\gamma}\eta.
\end{equation}

Here, $\gamma$ is the drag coefficient on a single bead, $kT$ is the thermal energy scale, $\eta$ is a delta-correlated normal random variable, and an overdot represents a time derivative. The final term provides a random force that emulates Brownian motion in a manner consistent with the fluctuation-dissipation theorem. The parameters of the system define its Lennard-Jones timescale, $\tau_{LJ}=\sigma\sqrt{m/\epsilon}$. These equations of motion are solved by LAMMPS \cite{lammps}, which iterates the system forward in time using the Velocity Verlet algorithm. We note that the foundation of our simulations is a LAMMPS tutorial found on Davide Michieletto's website \cite{davidewebsite} which was modified to study open and confined chains, and as such the description of the methods may be similar to previous works. 

We initialize torus knots with a harmonic parameterization scaled such that no bond is overstretched. If other knots are used, they are typically initialized from KnotPlot's coordinates \cite{scharein2002interactive} and rescaled to the desired number of beads with spline interpolation. To simulate untying, we initialize a knot in a closed chain and iterate it with using harmonic (rather than FENE) springs for 100 $\tau_{LJ}$ to thermalize the initial configurations and eliminate over- and under-stretching of the bonds, then another 100 $\tau_{LJ}$ with FENE bonds. This serves as an initial configuration for untying, after which the chain can be repeatedly untied by opening a single bond of the initial configuration, or randomly iterating through the bond that is opened. We typically iterate the opened chain for 1,000 to 5,000 100 $\tau_{LJ}$, depending on its size and knot complexity. Many of our simulations focus on the $7_1$ alternating torus knot, which unties through a predictable sequence of $7_{1}\rightarrow5_{1}\rightarrow3_{1}\rightarrow0_{1}$, (Figure 1b) each with an Alexander characteristic that is $\pm$ the crossing number. This untying sequence was also investigated by Caraglio et al. \cite{caraglio2019topological, caraglio2020topological}, and is similar to the even-twist untying sequence investigated by Soh et al \cite{soh2019conformational}.


\section{Results and Discussion}
\subsection{Validation and Performance}

\begin{figure}
    \centering
    \includegraphics[width=1\textwidth]{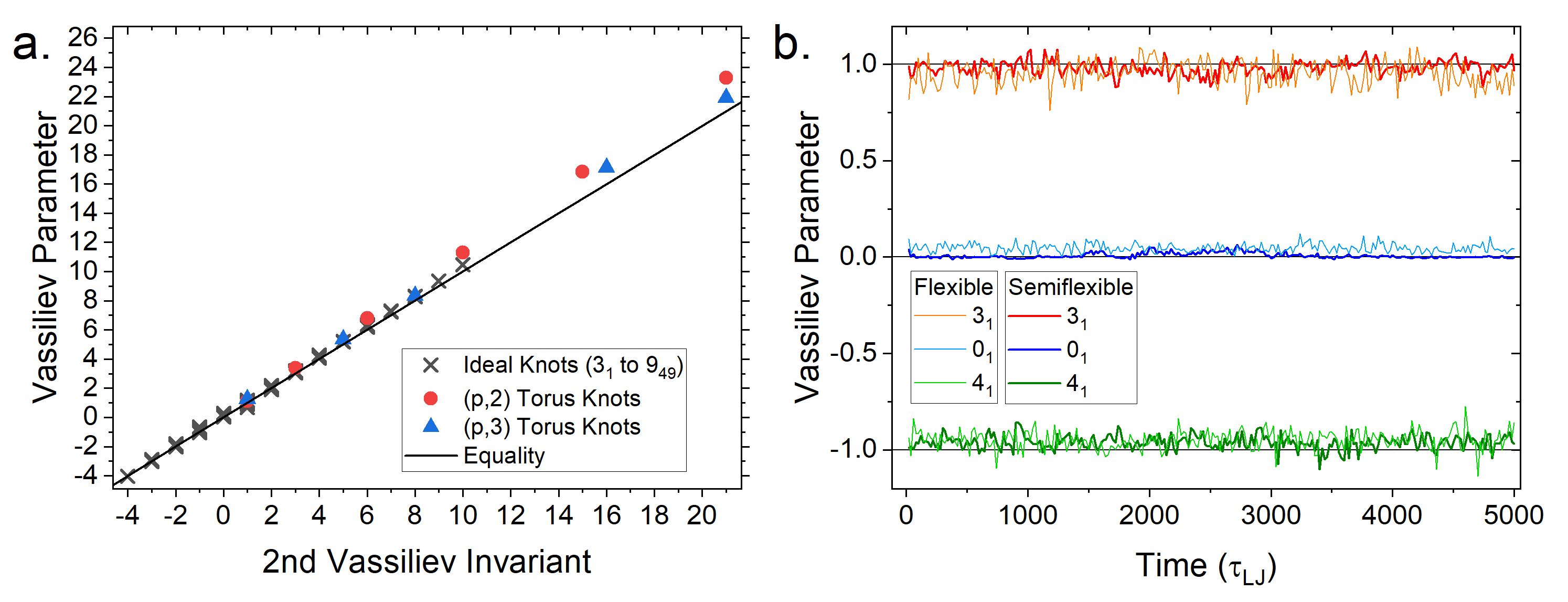}
    \caption{a. Scatter plot of measure Vassiliev parameter $V$ against the true value for ideal and torus knots. The solid line represents perfect agreement. b. Steady-state Vassiliev parameters for closed $0_1$, $3_1$, and $4_1$ knots simulated as polymers. The darker data curves represent semiflexible chains ($\ell_{p}=10\sigma$) and the lighter curves represent flexible chains ($\ell_{p}=1\sigma$). The expected values are shown as horizontal lines.}
    \label{fig:val}
\end{figure}

To verify that the Vassiliev parameter reproduces the Second Vassiliev Invariant for closed chains, we compute it for knots with smooth configurations of 256 vertices: ideal knots generated from Gilbert's coefficients for all knots up to 9 crossings \cite{gilbert}, and (p,2) and (p,3) torus knots. A plot of this validation is seen in Figure \ref{fig:val}a. Values of $v_2$ were taken from the coefficients of the Alexander-Conway polynomials on KnotInfo \cite{cha2011knotinfo}. Generally speaking, $V$ is close to but not identical to $v_2$, with the discrepancy increasing with $v_2$. Although there is a weak discrepancy between the predicted and measured values, for the knots sampled with nonzero $v_2$ the mean ratio between measurement and prediction was 1.01. We posit the disagreement lies in the discrete nature of our tangent vector calculation, as Ferrari and Zhao were able to more accurately measure $v_2$ with a corner-smoothing procedure \cite{ferrari2014monte}. To probe the worst possible case, we measured the Vassiliev parameter for a tight configuration of a 1023-crossing torus knot (presented in \cite{klotz2021ropelength}) with 400 vertices, an order of magnitude fewer vertices per crossing than typical of polymer knot simulations. Our algorithm underpredicted $v_2$ by 19\%.

To examine the utility of our algorithm for measuring the Vassiliev parameter for polymer knots, we simulated the steady-state behavior of closed polymers with $0_1$, $3_1$, and $4_1$ knots, at various persistence lengths. Fig. \ref{fig:val}b shows this data for $\ell_{p}=10\sigma$ and $\ell_{p}=1\sigma$. We expect the measurement to be less accurate and precise for flexible chains, when the tangent vector is not smooth and is poorly approximated by discretization. In both cases the measured Vassiliev parameter fluctuates near its expected value, and the flexible chain shows greater variance and a slightly larger offset from the expected value. Nevertheless, our discrete Vassiliev parameter approximates the second Vassiliev invariant quite well even for data it is not suited for, with the fluctuations and inaccuracies being much smaller than the difference between separate knot types. We also note that the fluctuations of the Vassiliev parameter are uncorrelated with typical measures of polymer knots. For example, for the semiflexible trefoil in Fig. \ref{fig:val}b, the radius of gyration and average crossing number are anticorrelated with a -0.64 Pearson's coefficient, while the Vassiliev parameter has a -0.16 and 0.09 correlation with the average crossing number and radius of gyration respectively. It is clear from Fig. \ref{fig:val}b that $V$ is not a perfect metric, but even in the worst cases its fluctuations and biases are small relative to the difference in $v_2$ between knots.

The algorithm we use was written in MATLAB and first calculates one half of the linking matrix (below the diagonal) in a two-nested for-loop, then computes SLL using a four-nested for-loop, requiring approximately $N^{4}/24$ multiplications of the matrix elements. We have included a MATLAB script in the ancillary data of this preprint. For comparison, we compute Alexander polynomial values with a MATLAB script that was adapted from one originally written by Benjamin Renner \cite{renner2015studying}. It projects the coordinates of a knot onto a surface and evaluates the over-underness of crossings to generate and computer the determinant of the Alexander matrix, similar to the KymoKnot algorithm \cite{tubiana2018kymoknot}. The simplest query of the Alexander polynomial will yield a value with indeterminate sign, which may be appropriate when the chain can only be in a few known knotted configurations (e.g. when untying between successive (p,2) torus knots). Multiple queries (in our case, four) are required to determine the sign, and projecting the endpoints of the knot to multiple locations on an enclosing sphere will require significantly more. 

\begin{figure}
    \centering
    \includegraphics[width=0.6\textwidth]{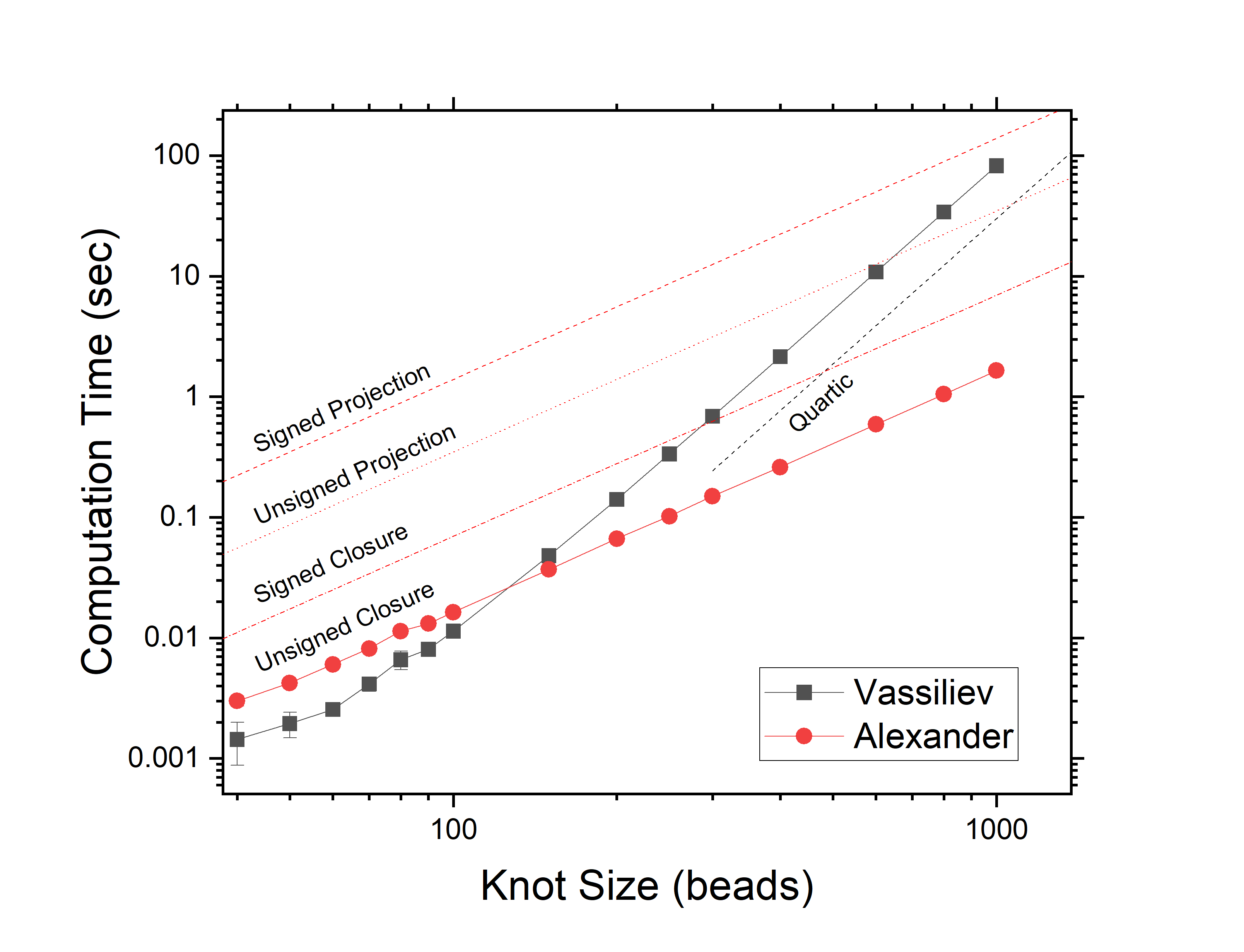}
    \caption{Mean computation time required to compute the Vassiliev parameter and the Alexander characteristic of four knots as a function of their size. Parallel dashed lines approximate the Alexander computation if sign determination and sphere projection is used. Error bars, when visible, represent the standard error on the mean, and are smaller than the symbols otherwise.}
    \label{fig:perf}
\end{figure}

We calculated the computation time required to calculated $V$ for the unknot, $3_1$, $5_1$, and $7_1$ knots for simple configurations with between 40 and 1000 vertices. We compared this to the computation time of the unsigned Alexander characteristic ($\Delta(-1)$), which we multiply by 4 to take into account sign determination, and 20 to take into account projection to uniform points on a sphere. The results are seen in Figure \ref{fig:perf}. This computation was performed in 2023 on a Lenovo laptop with a 3 GHz AMD Ryzen 5 4600 H processor and 16 GB of RAM. It is entirely possible that more efficient algorithms, more powerful hardware, or GPU optimization could reduce the computation time of either parameter. 

The Vassiliev computation is asymptotically quartic, whereas the simplest Alexander computation is asymptotically quadratic, and will be faster for large knots. However, the Vassiliev calculation is faster than the Alexander calculation below approximately 120 vertices. Requiring more complexity out of the Alexander computation increases the size at which it begins to outperform the Vassiliev computation: approximately 300 for the signed Alexander characteristic, 600 for an unsigned projection to 20 points, and well over 1000 for a signed projection. Spherical projection typically uses more than 20 closure points, which would extend V's competitive range. The 40-bead knots take 1-2 milliseconds, the largest about 82 seconds. In practice, an untying simulation of an open knotted chain with 100 monomers could be analyzed in about two seconds, depending on the sampling density. We believe $V$ satisfies our fifth requirement, that of speedy computation, for all but the largest knots, and is competitive with the Alexander polynomial especially if multiple closure points are desired. In principle, the time required to compute the Alexander polynomial increases with knot complexity, as the determinant of a larger crossing matrix must be taken. In practice, this did not have a significant effect on the computation time until the tested knots had 50 or more essential crossings, and at N=400 exceeded the Vassiliev computation time at around 67 crossings. This may be relevant for highly confined or very long chains, although in the latter case the Vassiliev measure would likely still be slower.

Another freely available package, Knoto-ID \cite{dorier2018knoto}, can efficiently characterize open curves as projection-dependent distributions of knotoid diagrams, and may satisfy the five requirements listed in the introduction. A potential disadvantage of knotoids is combinatorial explosion: there are 2 knots with 5 crossings, and 950 knotoids. Categorization of higher-crossing knotoids is incomplete, and the interpretation of a partially untied knot being consistent with dozens of knotoids rather than two knots is at this time ambiguous. While we did not extensively test it, we found that Knoto-ID was able to characterize the knotoids of an N=100 $7_1$ knot untying over 200 time points, along 20 projected axes in about five seconds. The Vassiliev parameters for the same data set were computed in 2 seconds, but would not compare as favorably for longer chains. A full evaluation of Knoto-ID for characterizing confined and untying polymer knots is beyond the scope of this work.

\subsection{Untying Polymers}

\begin{figure}
    \centering
    \includegraphics[width=0.99\textwidth]{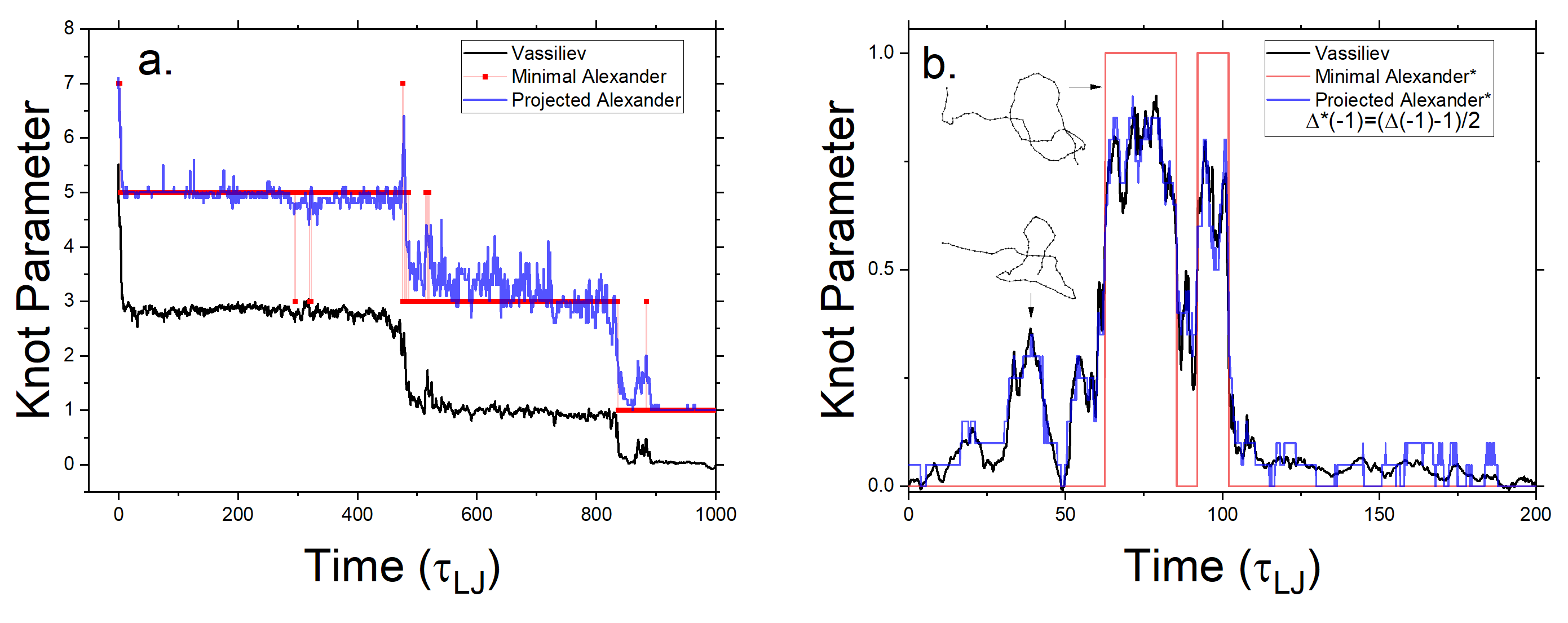}
    \caption{Time series of the Vassiliev and absolute Alexander parameters of knots as they untie and retie. a. A knot initialized as a $7_1$ and untying through a sequence of $5_1$ to $3_1$ to the unknot. The expected values are 7-5-3-1 for the absolute Alexander polynomial and 6-3-1-0 for the Vassiliev parameter. The red curve shows the Alexander parameter as determined by minimally interfering closure, and the blue shows the average value when the ends are projected to 20 evenly spaced points on a large enclosing sphere. b. The Vassiliev and Alexander parameters during an event in which an unknotted polymer spontaneously forms a trefoil knot. In this case, the Alexander value has been rescaled so that it is easier to compare it to the Vassiliev parameter by eye.}
    \label{fig:untiejumps}
\end{figure}

Here we discuss the use of the Vassiliev parameter for the analysis of unconfined knotted polymers. Our simulations typically initialize a closed knotted chain with $N=100$ beads and then open the chain, simulating it for long enough that it is unambiguously unknotted. An example is seen in Figure \ref{fig:untiejumps}a, in which the absolute Alexander characteristic and the Vassiliev parameter are plot for a chain that is initialized as a $7_1$ and unties to the unknot. Figure \ref{fig:untiejumps}b shows the case of an unknotted chain transiently forming a trefoil knot and untying again. While Fig.\ref{fig:untiejumps}a shows the results of a single simulation, the population average of 50 untying $7_1$ knots is shown in Fig. \ref{fig:sphere}.

To better compare the Alexander polynomial to the Vassiliev parameter, we define the \textit{averaged absolute Alexander characteristic} which is the mean value of the absolute value of the Alexander polynomial evaluated at -1 with the ends of the knot closed at 20 different points corresponding to the vertices of a large regular dodecahedron surrounding the chain. For strongly knotted chains, this has a similar value to the Alexander polynomial evaluated through minimally interfering closure, while for weakly knotted chains it represents the many possible knots consistent with the chain. We take the absolute value to better show-correlation between the averaged Alexander characteristic and the Vassiliev parameter, as the Alexander characteristic of torus knots has a varying sign while $v_2$ is strictly positive. To better demonstrate the correlation between this parameter and $V$, we can transform the Alexander polynomial such that $\Delta*=(|\Delta|-1)/2$ so that it takes values of 1 and 0 for the trefoil and unknot, the same as the Vassiliev values. 

The single-closure Alexander polynomial is an excellent tool for detecting and characterizing knots when the topology is stable and the chain ends are far from the knot. When the knot is in the process of untying and one or both of the chain ends are within the core of the knot, it can produce wildly fluctuating results. Typically, the Alexander value will jump between (at least) two values as one knot transitions to another or the unknot. An advantage of the Vassiliev parameter is that is insensitive to the choice of closure and does not display these jumps. For example, Fig. \ref{fig:untiejumps}a shows the untying of a $7_1$ knot as determined by both parameters. The $7_1$ quickly becomes a $5_1$ but the Alexander characteristic fluctuates between its values of 7, 5, and 3 as closure identifies it as a $7_1$, $5_1$, or $3_1$, before it enjoys brief stability at 3 before repeating the same process between 3 and 1, its unknotted value. In contrast, the Vassiliev parameter undergoes small fluctuations, but never undergoes ambiguous flips in its knot type. It decreases through its values of the four knots, 6 to 3 to 1 to 0 without reversal.


In Figure \ref{fig:untiejumps}b the rescaled sphere-averaged Alexander polynomial and the Vassiliev parameter are seen to be highly correlated, indicating they contain similar information about the topological state of the polymer. Both cases show several features that highlight the usefulness of the Vassiliev parameter compared to the single-closure Alexander polynomial. The minimally interfering Alexander parameter shows discrete jumps when the knot type is determined to have changed, whereas the Vassiliev parameter smoothly evolves from one state to another. 
Although $V$ fluctuates when the knot is steady and $\Delta$ does not, the fluctuations contain topological information (e.g. the rise at around t=40 in Fig. \ref{fig:untiejumps}b showing that the chain is in a position where it could soon tie, but doesn't). In longer simulations of dynamic knotting and unknotting (e.g. of a strongly confined polymer chain), the autocorrelation function of $V$ will provide information about topological dynamics in a way that $\Delta$ is unable to.

\begin{figure}
    \centering
    \includegraphics[width=0.6\textwidth]{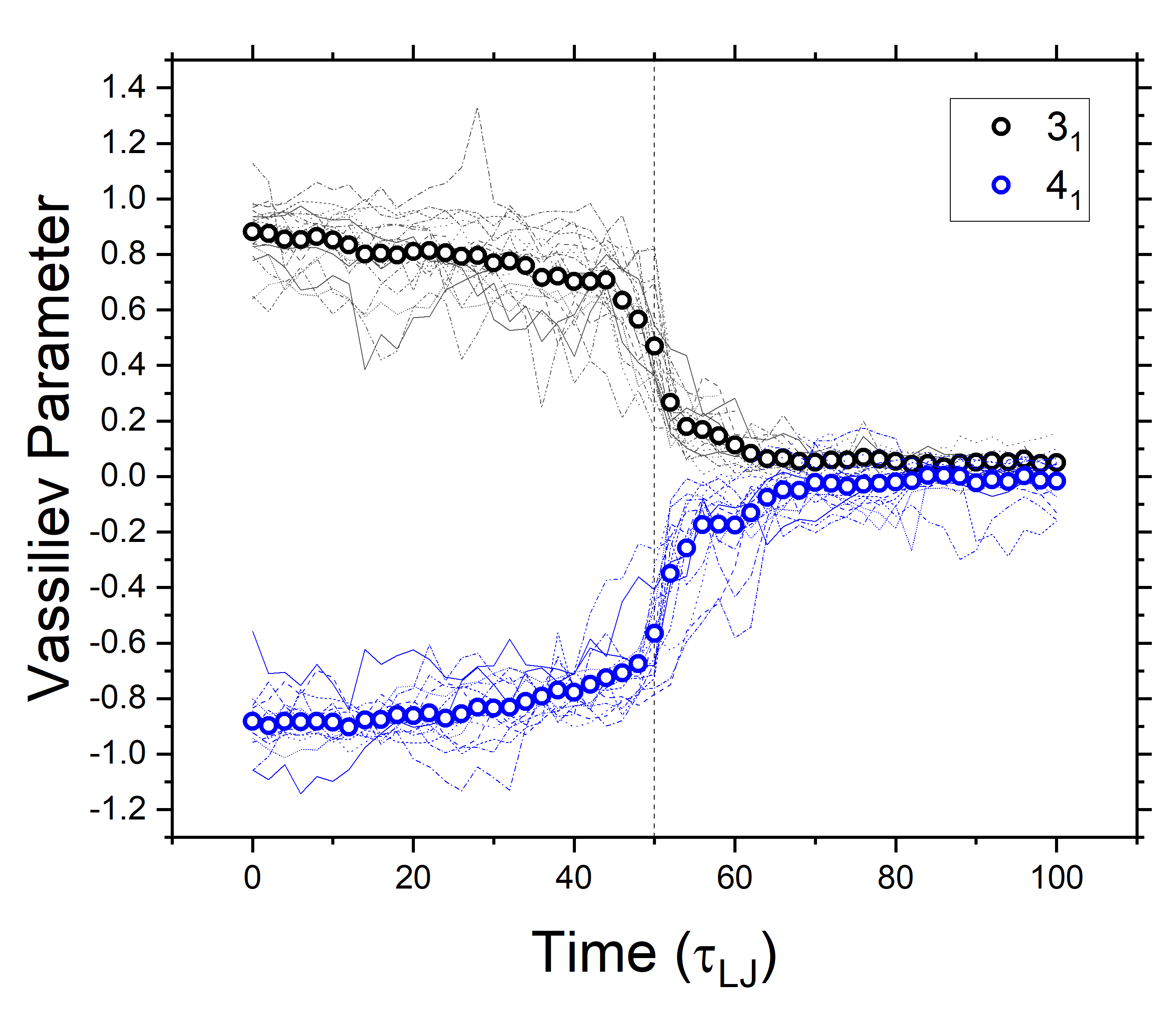}
    \caption{Vassiliev parameter of populations of $3_1$ and $4_1$ polymer knots untying, aligned at the time at which their minimally-closed Alexander polynomial shifts (vertical line). The population average is overlaid.}
    \label{fig:untiealign}
\end{figure}

The closure-Alexander method and other discrete classifiers \cite{sleiman2024geometric} can detect the point of transition between two knot types on an untying pathway or between a knot and the unknot. However, this is typically not a discrete process, as the polymer undergoes conformational changes in preparation for untying (for knotted DNA in elongational fields, this can include a contraction of the chain \cite{soh2018untying}), and immediately after still maintains vestiges of its former knot. Previous studies have examined the survival probability function of ensembles of knots \cite{caraglio2019topological}, but this loses information about the knottedness during the untying event. To closely examine the polymer untying process, we simulated $3_1$ and $4_1$ knots untying to the unknot, and measured $V$ going from $\pm1$ to 0. While these knots untied at a stochastic range of times, we measured the minimally-closed Alexander polynomial at -1 and identified the time at which it shifted. We used this time to align the Vassiliev data from many runs, and examined its population average during the untying event (Fig. \ref{fig:untiealign}). Both knots have similar untying trajectories, with the average curve showing three phases. First, the absolute Vassiliev parameter slowly decreases below its knotted value. At the same time when the Alexander polynomial shifts, it makes a substantial jump towards zero, then further levels off towards zero. A comparable plot using the Alexander polynomial would reveal a step function. This suggests the Vassiliev parameter as a new tool for studying the polymer untying process, and its behavior with respect to the physical parameters of the system (e.g. chain length) is an intriguing subject for future work.

\subsection{Confined Polymers}

A knotted polymer confined in a sphere typically has ends that cannot be closed without introducing a significant number of new crossings, leaving the knot type indeterminate under traditional methods. A spherical closure scheme will produce a distribution of knot types for a typical configuration, which will have a mode that makes up less than 50\% of total angles, putting confined polymers in the weakly knotted regime.

\begin{figure}
    \centering
    \includegraphics[width=\textwidth]{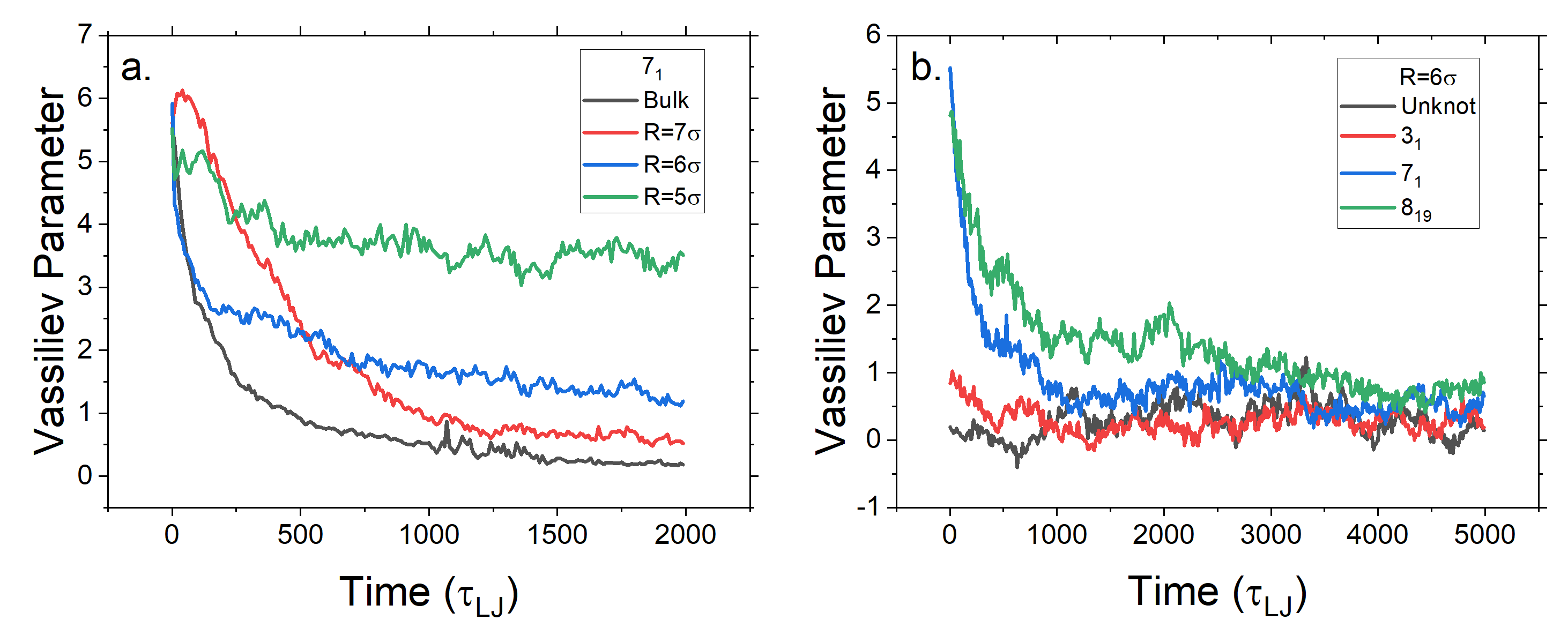}
    \caption{a. Population-average Vassiliev parameter of ensembles of $7_1$ knots untying under varying degrees of confinement. An equilibrium level is reached in each case, which increases as the confinement becomes tighter. b. Population-average Vassiliev parameter of ensembles of $0_1$, $3_1$, $7_1$ and $8_{19}$ knots untying in a sphere of R=6$\sigma$. They eventually reach a common equilibrium level of the Vassiliev parameter.}
    \label{fig:sphere}
\end{figure}

Figure \ref{fig:sphere}a shows the evolution of an ensemble average $V$ of initialized $7_1$ knots evolving in confining spheres of different radius. In the bulk, this decays to zero as the knot unties, similar to the knot survival probability analyzed by Caraglio et al. \cite{caraglio2019topological}. Under confinement however $V$ reaches a plateau, which can be imagined as a chemical equilibrium between knot untying and the formation of new knots. The average value of this plateau increases with the strength of confinement. Figure \ref{fig:sphere}b. shows the evolution of several initial knots under the same confinement. While they start at their expected value, they approach the same plateau at approximately the same rate, indicating a timescale beyond which the initial topology is no longer relevant. 

\begin{figure}
    \centering
    \includegraphics[width=0.7\textwidth]{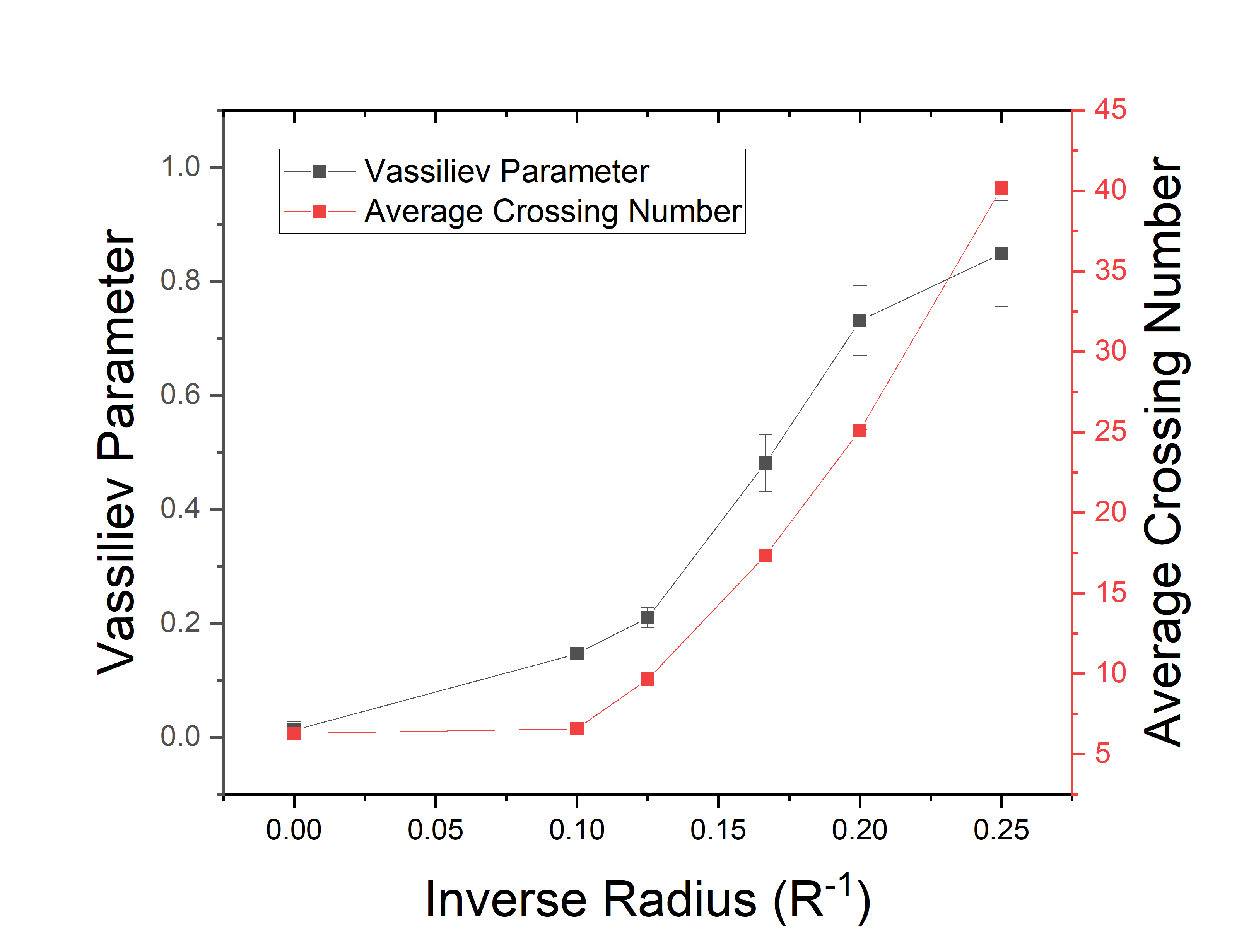}
    \caption{Ensemble average Vassiliev parameter and average crossing number for populations of 20 initially unknotted polymers confined in spheres and in the bulk, as a function of the reciprocal confinement radius.}
    \label{fig:SphereVals}
\end{figure}

Treating $V$ as a general measure of the knottedness of confined chains that isn't necessarily identified with a single knot invites comparison to the average crossing number, which can be computed more efficiently. The ACN has previously been used to characterize the knottedness of spherically confined polymers to study viral capsid packaging \cite{arsuaga2009growth}. $V$ is more closely tied to topology than ACN, as ACN is strongly anti-correlated with radius of gyration and is sensitive to nugatory folding, while $V$ is not. However, an increasing ACN in confinement generally indicates stronger knotting, which is not necessarily the case for $V$. For example, an $8_{14}$ knot could undergo strand passage into a $3_1$ knot, and $V$ would go from 0 to +1. Similar events in spherical confinement mean that the sign of $V$ fluctuations are not necessarily correlated with more complexity. Although values may take either sign, the data indicates that $V$ is increased by confinement, which is consistent with the findings of Smith and Panagiotou for random walks confined in spheres \cite{smith2022second}. Fig. \ref{fig:SphereVals} shows the population average value of the Vassiliev parameter and the average crossing number for 20 initially unknotted polymers in confinement (R=4 to R=10) and the bulk, showing that both parameters increase similarly in confinement. While Smith and Panagiotou have proven the positivity of these values, the interpretation of these specific numbers is an open question.

\subsection{Knotted Subchain Identification}

One use of knot identification algorithms is to locate the knotted subchain on a larger polymer. This has been useful for finding the equilibrium sizes of knots in closed chains to validate models of knot metastability \cite{dai2015metastable}, as well as identifying configurations on open chains that lead to different untying mechanisms while tracking the motion of knots \cite{soh2019conformational}. Typically this is done by iteratively removing beads from the end of the chain and calculating the Alexander polynomial at each stage, noting at which point the topology of the chain changes (it may also be done ``bottom-up'' by adding beads to a virtual subchain until a knot is detected). This can produce a discrete step function of knottedness with respect to monomer index. Recently, Barbensi and Celoria extended this to describe the Knot Intensity Distribution, which quantifies the importance of an individual monomer to the global state of the knot \cite{barbensi2022knot}. A similar scheme can be performed with the Vassiliev parameter, except is does not produce a binary clasification of knottedness, but rather a local measure of salience that reflects how much $V$ changes when a subchain terminating in that bead is removed. This can fill a similar role as the Alexander subchain detection, with a slightly different interpretation.

\begin{figure}
    \centering
    \includegraphics[width=\textwidth]{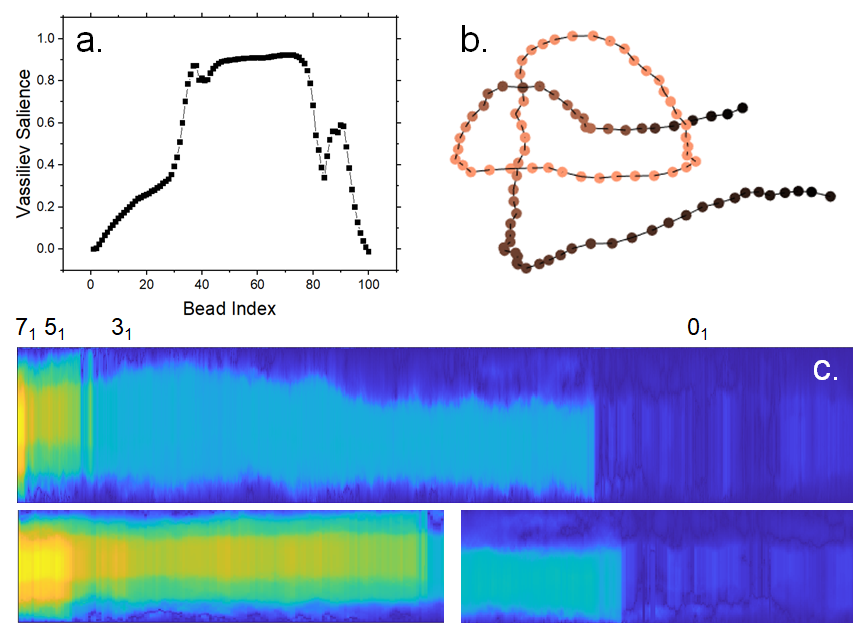}
    \caption{Knotted subchain analysis using the Vassiliev parameter. a. Vassiliev salience, the degree to which a subchain between each monomer and the closest end contributes to the total Vassiliev parameter for an open trefoil knot. b. Conformation of the knot analyzed in a., with brighter monomers corresponding to greater salience. c. Kymograph of the Vassiliev salience for an initial $7_1$ knot untying towards the unknot in several stages (including a brief retying from $3_1$ to $5_1$), from which the trefoil in a. and b. was sampled. The smaller kymographs below show the initial untying from $7_1$ to $5_1$ to $3_1$, and the final untying from $3_1$ to the unknot.}
    \label{fig:kymo}
\end{figure}

We calculate a ``top down'' measure of the salience of each monomer by removing monomers from the end until the Nth bead and calculating $V$. We subtract that from the full chain's Vassiliev parameter to find a measure of how much $V$ would change if the subchain from that bead to the closest end were removed. An example for a trefoil knot can be seen in Figure \ref{fig:kymo}a-b. This can also be applied to closed chains with an arbitrary cut point, and is insensitive to the choice of starting point. There are a few differences between an Alexander-based subchain detection and the Vassiliev salience measure. The value within the knotted core is not constant, but typically reaches a plateau. There is also no fixed cutoff beyond which the chain is definitively not part of the knot. There is a drop-off towards the end of the chain, as the Vassiliev parameter depends more on the orientation of the free ends than does the Alexander polynomial. The Vassiliev salience at each stage of a knot untying sequence can be extended into a kymograph (Figure \ref{fig:kymo}c), similar to those shown by Caraglio et al. \cite{caraglio2019topological} or Tubiana et al. \cite{tubiana2013spontaneous}. When the knot is still consistent with $7_1$ or $5_1$, multiple levels within the knotted section of the kymograph can be seen. The unknotted portion of the kymograph contains a ``ghost'' region of where a knot would form if a chain end were to penetrate it. While we have considered only the detection of knotted subchains, in principle a subchain matrix between all pairs of monomers could be calculated, and used to identify slipknots and pre-knotted segments of the chain.

\section{Concluding Remarks}

We believe the analyses present demonstrate that the Vassiliev parameter is a useful tool for analyzing knot simulations, providing additional information not revealed by a single-closure Alexander polynomial measurement. However, for sufficiently long chains, it will be considerably slower. A more direct comparison would be to stochastic spherical closure, to which the Vassiliev parameter compares favorably. For the analysis of closed chains or the validation of topology in a simulation in which pathological strand-crossing may occur, the Alexander polynomial is likely superior. 

The algorithm discussed in this manuscript is sufficient for chains that are up to a few hundred beads long, consistent with typical lengths used to study knotted polymer dynamics. Sleiman et al. \cite{sleiman2024geometric} developed a new knot classification algorithm based on chains of N=100, Caraglio et al. \cite{caraglio2019topological} simulated untying knots in chains of up to N=200, while Soh et al. \cite{soh2019conformational} used N=300. Tubiana et al. simulated knotting and unknotting for chains of up to N=4096, which is likely beyond the useful range of the current algorithm. Likewise, studies of knot formation in growing static polymers typically use much longer chains, for example up to 40,000 beads by Rieger et al. \cite{rieger2016monte}, for which our algorithm is not useful.

Recently, Sleiman et al. \cite{sleiman2024geometric} found that a neural network trained on polymer knots in a ``local writhe'' basis could effectively categorize and identify knots, including during untying. The writhe basis used by Sleiman et al. generates a matrix for each configuration that is similar to the linking matrix in Equation \ref{eq:matrix}, except that each element categorizes how the local chain around monomer i is linked with the local chain around monomer j. We can compute a version of the Vassiliev parameter based on the local segment-to-segment writhe matrix, and find that it yields similar results. Comparing the Vassiliev parameters calculated from a) the regular writhe matrix to those from b) the segment-to-segment writhe matrix, using the same polymer model as Sleiman et al., yields values of 0.003 and 0.007 for the unknot ($v_2$=0), 0.98 and 0.94 for the trefoil ($v_2$=1), and 2.92 and 2.82 for the $5_2$ ($v_2$=3). Because Sleiman et al.'s neural network effectively performs a matrix convolution of a write matrix, the training algorithm may be discovering something similar to (but stronger than) the Vassiliev invariant.

Overall, we have demonstrated that the Vassiliev paremeter, a discretized version of the Second Vassiliev Invariant for open curves, is a useful tool for characterizing open polymer knots in conditions where the topology is ambiguous, including during untying and under strong confinement. The algorithm can be computed from Cartesian coordinates, produces values consistent with those expected for closed knots when appropriate, interpolates between these known values during untying, can identify the salient portions of knotted subchains, and performs these computations in a reasonable time. Additionally, it identifies several features not observed using a single-closure Alexander measure, such as fluctuations associated with unsuccessful knotting events, the precursors and postcursors of untying, and a confinement-dependent average value. Its use having been established in this work, future studies can investigate deeper aspects of the physics of polymer untying and knot formation using the Vassiliev parameter.

\section{Acknowledgements}
This work was supported by the National Science Foundation, award number 2105113. We are appreciative of feedback from Eleni Panagiotou and Cristian Micheletti. We thank Davide Michieletto for providing an online LAMMPS tutorial, but have not discussed this work with him.

\bibliographystyle{unsrt}
\bibliography{knotrefs}

\end{document}